\documentclass[a4paper,11pt]{article}
\usepackage{aaskaiid}
\usepackage{multirow}
\usepackage{orcidlink}

\newcommand{\aastar}{AA$^{*}$}

\newcommand{\gmrt}{uGMRT}
\newcommand{\vla}{VLA}
\newcommand{\lofar}{LOFAR}

\newcommand{\fesc}{$f_\mathrm{esc}^\mathrm{LyC}$}
\newcommand{\alphanth}{\ifmmode \alpha_\mathrm{nth} \else $\alpha_\mathrm{nth}$\fi}
\newcommand{\alphacs}{$\alpha^{\mathrm{3GHz}}_\mathrm{6GHz}$}
\newcommand{\Oratio}{O$_{32}$}

\newcommand{\hii}{H{\sc ii}}

\newcommand{\fth}{\ifmmode \mathrm{f}_{\nu} ^\mathrm{th} \else f$_{ \nu }^\mathrm{th}$\fi}
\newcommand{\FHbeta}{\ifmmode \mathrm{F}^{\mathrm{dustcorr}}_{\mathrm{H}\beta} \else $\mathrm{F}^{\mathrm{dustcorr}}_{\mathrm{H}\beta}$ \fi}
\newcommand{\sth}{\ifmmode S_{\mathrm{1GHz}}^{\mathrm{th}} \else S$_{\mathrm{1GHz}}^{\mathrm{th}}$\fi}
\newcommand{\snth}{\ifmmode S_{\mathrm{1GHz}}^{\mathrm{nth}} \else S$_{\mathrm{1GHz}}^{\mathrm{nth}}$\fi}
\newcommand{\nuturn}{\ifmmode \nu_\mathrm{t} \else $\nu_\mathrm{t}$\fi}

\newcommand{\mstar}{$\mathrm{M_*}/\mathrm{M_\odot}$}
\newcommand{\rcsed}[1]{radio-SED{#1}}

\newcommand{\sbs}{SBS 0335-052}
\newcommand{\izw}{I Zw 18}

\defcitealias{Hunt04}{H04}
\defcitealias{Sebastian19}{SB19}

\newcommand{\revtext}[1]{#1} 

\title{Probing the Nature of Lyman Continuum Emitting and Low-metallicity Galaxies Using the SKA}
\ShortTitle{SKA observations of LyC leakers and metal-poor galaxies}

\author[1, 2]{Omkar Bait\orcidlink{0000-0003-2722-8841}}
\ShortName{Bait et al.} % shortened name list for header 
\author[3,4]{Daniel Schaerer\orcidlink{0000-0001-7144-7182}}
\author[5]{Mark Sargent\orcidlink{0000-0003-1033-9684}}

\affiliation[1]{National Radio Astronomy Observatory, 520 Edgemont Road, Charlottesville, VA 22903, USA}
\affiliation[2]{The NSF-Simons AI Institute for Cosmic Origins, USA, 201 E. 24th Street, POB 4.102, Austin, Texas 78712-1229} 
\emailAdd{obait@nrao.edu}
\affiliation[3]{Observatoire de Gen\`eve, Universit\'e de Gen\`eve, Chemin Pegasi, 1290 Versoix, Switzerland }
\emailAdd{daniel.schaerer@unige.ch}
\affiliation[4]{CNRS, IRAP, 14 Avenue E. Belin, 31400 Toulouse, France }
\affiliation[5]{Institute of Physics, Laboratory of Astrophysics, Ecole Polytechnique Federale de Lausanne (EPFL), Observatoire de Sauverny, CH-1290 Versoix, Switzerland}
\emailAdd{mark.sargent@epfl.ch}

\abstract{
\revtext{The sources responsible for cosmic reionization remain a key open question in observational cosmology. Recent \textit{JWST} results increasingly suggest that low-mass star-forming galaxies (e.g., compact starbursts and strong emission-line systems) dominated the ionizing photon budget. The physical mechanisms driving Lyman continuum (LyC) photon escape, including supernova, radiative, and cosmic-ray feedback, and the origin of extreme ionization conditions, remain poorly understood. Radio continuum (RC) emission, a well-established star-formation tracer in normal galaxies, is not yet well characterized in such extreme systems, which exhibit high star-formation rate densities, young stellar populations, low metallicity, and hard ionizing spectra.}

\revtext{Targeted mid-frequency (1--15\,GHz) observations with SKA precursors have begun probing low-redshift LyC emitters (LCEs), revealing links between RC spectral index, LyC escape fraction, ionization conditions, metallicity, and SFR surface density, alongwith deviations from the canonical RC--SFR relation. The higher sensitivity of the SKA Array Assemblies across Bands~1--5 will enable systematic studies of fainter LCEs and low-mass, metal-poor galaxies. We present number density predictions for LCE candidates at $z \sim 1$--$3$, showing that SKA-Mid surveys can assemble samples of $\sim$10--100 candidates per square degree over a star-formation rate range of 1--100\,$M_{\odot}$\,yr$^{-1}$, making a dedicated SKA Large Programme scientifically feasible.}

\revtext{With the full SKA, in synergy with \textit{JWST} and next generation telescopes, multi-wavelength analyses will robustly constrain the thermal and non-thermal RC components and cosmic-ray energy spectra, providing critical insights into the feedback processes governing LyC escape and star formation in the early universe.}
}

\begin{document}
\maketitle

\section{Introduction}

Understanding how the Universe transitioned from a neutral to a fully ionized state remains one of the central challenges in modern astrophysics. Over the past decade, major advances in optical and ultraviolet \revtext{facilities} have revealed that star-forming galaxies likely played the dominant role in reionizing the intergalactic medium (IGM). Yet, the exact nature of these sources and the physical mechanisms that allowed Lyman continuum (LyC) photons to escape remain only partially understood. The fraction of ionizing photons that successfully escape into the IGM, commonly referred to as the escape fraction (\fesc{}), is a key parameter in quantifying the contribution of galaxies to cosmic reionization. Constraining \fesc{} and understanding the conditions that regulate it are therefore fundamental to connecting the observed galaxy population with the ionization history of the early Universe.

The \textit{James Webb Space Telescope} (JWST) has recently provided a wealth of insights into the population of faint, low-mass galaxies at high redshift ($z \geq 6$), revealing compact starbursts with intense emission-line spectra and very low metallicities \citep[e.g.,][]{Schaerer22, Rhoads23, Langeroodi23, Topping24}. These galaxies possess a high ionizing photon production rate \citep[e.g.,][]{Simmonds23, Simmonds24, Saxena24,Atek24} and could potentially be LyC leakers. To better understand these distant systems, researchers have turned to their low-redshift analogues, where detailed multi-wavelength studies are feasible with current-generation facilities.

Some of the well-known examples of local analogues of high-$z$ galaxies in the literature are I Zw18 \citep{IZW18-localanologue}, SBS 0335-052 \citep{Izotov09-SBS, Hunt04}, Haro 11 \citep{Haro1106-LyCescape}. \revtext{Over time}, a growing class of compact, metal-poor star-forming galaxies characterized by high [OIII]/[OII] ratios (\Oratio{}), strong nebular emission lines, and extreme ionization parameters has emerged as the most promising local analogue of reionization-era galaxies \citep{Izotov11, Izotov21a-CSFGs, Schaerer16, Izotov21b, Izotov24-LyA-metalpoor}. These Extreme Emission Line Galaxies (EELGs) are referred by various names in the literature e.g., Green Peas \citep[GPs;][]{cardamone2009, Izotov11-GPs}, blueberries \citep[BBs;][]{yang2017, Kouroumpatzakis24_BB}, extremely metal deficient dwarfs \citep[XMDs; e.g.,][]{Izotov07-XMD, Pustilnik11-XMD,  Izotov12_XMD, Berg12-XMD, Guseva15-XMD}. 

Direct LyC leakage was previously known in only a handful of EELGs \citep[e.g.,][]{Haro1106-LyCescape, Borthakur14}. In a series of pioneering studies Yuri Izotov and collaborators identified several EELGs at low-$z$ with strong LyC leakage, revealing \fesc{} ranging from a few percent up to nearly 50\% \citep{Izotov16a, Izotov16b, Izotov18a, Izotov18b-vsep, Izotov21b}. These galaxies tend to be low-mass (\mstar{} $\leq 10^{9}$~M$_\odot$), low-metallicity (Z~$\sim$~0.1--0.2~Z$_\odot$), and extremely compact, often hosting intense bursts of star formation. This breakthrough motivated the Low-$z$ Lyman Continuum Survey (LzLCS) to conduct a systematic study of LyC leakers (LCEs) at low-$z$, $z \sim 0.3$ \citep{Flury22a, Flury22b}. LzLCS provides a statistically robust sample of low-$z$ LCEs using the \textit{Hubble Space Telescope} (HST) Cosmic Origins Spectrograph (COS). The LzLCS has uncovered empirical correlations between the LyC escape fraction and several galaxy properties, including \Oratio{}, star-formation surface density, and outflow kinematics, highlighting the importance of stellar feedback and low neutral-gas covering fractions in enabling LyC escape. A common finding from these studies is that the strongest LCEs generally have very compact sizes, very high star-formation-rate (SFR) surface densities, very high ionization ratios (\Oratio{}), very young stellar populations, very low dust content, and extremely low metallicities. The LzLCS and related LCE samples provides a crucial local benchmark for interpreting the reionization-era sources now being discovered with JWST.

Throughout this chapter, we refer to these galaxies as LCEs which will be the main focus of this chapter. \revtext{We stress, however, that the currently confirmed sample of LCEs which are
galaxies with directly detected LyC emission remains small and is 
largely restricted to $z \lesssim 0.3$, where space-based UV 
spectroscopy is feasible \citep{Izotov16a, Izotov16b, Izotov18a, 
Izotov18b-vsep, Izotov21b, Flury22a}. At higher redshifts, direct 
LyC detection becomes inaccessible due to increasing IGM opacity, and 
one must rely on indirect indicators, e.g., extreme rest-frame 
EW$_{H\alpha}$, high \Oratio{}, compact morphology, and low dust 
content to identify galaxies that are likely, but not confirmed, 
LyC leakers. Throughout this chapter, we therefore use the term 
\textit{LCE candidates} when referring to EELGs selected by such 
indirect criteria, reserving \textit{confirmed LCEs} for the 
low-redshift sample with direct LyC detections. The question of how 
one identifies and confirms LCEs in practice in a large radio-selected 
sample through, e.g., UV spectroscopy, emission-line 
diagnostics, or resolved morphology is an important one but lies 
beyond the scope of this chapter.}

Despite these advances, the radio continuum (RC) properties of LCEs remain poorly explored. RC emission arising from both thermal free–free processes and non-thermal synchrotron radiation, provides a dust-unbiased probe of star formation, supernova feedback, cosmic-ray populations, and magnetic fields \citep{condon1992, murphy2011}. Recent targeted radio studies using SKA precursors and pathfinders such as the Jansky Very Large Array (JVLA), the upgraded Giant Metrewave Radio Telescope (uGMRT), and the Low-Frequency Array (LOFAR) have begun to reveal intriguing connections between the radio spectral index, metallicity, star-formation rate surface density, and \fesc{} in low-$z$ LCEs \citep{Bait24a}. These results suggest that radio continuum observations can serve as an independent and physically motivated tracer of the mechanisms that regulate LyC escape. Multi-frequency \rcsed{} studies of nearby LCEs have shown that these systems often exhibit atypical radio spectra not observed in normal star-forming galaxies \citep{Bait25}.

The \textit{Square Kilometre Array} \citep[][SKA]{braun2019anticipatedperformancesquarekilometre} will provide a transformative step forward in this field. Its unprecedented sensitivity, broad frequency coverage, and high angular resolution in the mid-frequency range (1–15~GHz) will enable systematic radio studies of fainter and more representative samples of low-$z$ LCEs, probing lower stellar masses and metallicities than has been possible to date. Equally importantly, this science driver offers a natural synergy with the SKA-LOW array. While SKA-Mid will directly constrain the physical conditions and feedback processes governing LyC escape in galaxies, SKA-Low will trace the large-scale distribution of ionized and neutral hydrogen during the Epoch of Reionization (EoR) \citep{Koopmans15-SKALOW-EoR}. Together, these observations will connect the properties of ionizing sources to the evolving ionization structure of the early Universe, providing a unified framework linking local LCEs, their high-redshift analogues, and the reionization history itself.

This chapter focuses on the prospects of detecting a wider population of LCEs and EELGs in the radio using the SKA-Mid bands and is organized as follows. In Section~\ref{sec: physical properties}, we describe the nature of feedback in LyC-emitting galaxies. Section~\ref{sec: ska precursors results} summarizes recent radio studies of these systems using existing SKA precursors. In Section~\ref{sec: ska prediction}, we discuss the prospects for detecting and characterizing these galaxies at low-$z$ with the SKA-Mid AA$^*$ and AA4 configurations. \revtext{In Section \ref{sec: LCE high-z} we discuss the prospects to detecting candidate LyC emitting and metal-poor galaxies at high redshift using various SKA extragalactic continuum reference survey scenarios. }Finally, we summarize the main conclusions in Section~\ref{sec: conclusions}.

\section{The nature of feedback in LyC emitting galaxies}
\label{sec: physical properties}

Feedback plays a key role in shaping the geometry and porosity of the ISM, thereby facilitating the escape of LyC photons. In particular, SN feedback can redistribute and disperse the gas, creating regions that become optically thin to LyC photons while others remain optically thick, resulting in a picket-fence morphology \citep[e.g.,][]{Jaskot19}. In some cases, strong SN feedback can drive large-scale outflows and open holes or chimneys in the ISM, through which LyC photons can efficiently escape \citep[e.g.,][]{Heckman11}. On the other hand, in low-metallicity environments dominated by young, massive stars, ionization feedback may play an equally or even more important role than SN feedback in maintaining the ISM in a highly ionized state, thereby enhancing the LyC escape \citep[e.g.,][]{Gazagnes20}.

In the framework of cosmological and zoom-in simulations, SN feedback has often been identified as the dominant channel through which LyC photons leak from galaxies \citep[e.g.,][]{Kimm14, Paardekooper15, Trebitsch17, Ma20}. In these models, LyC escape is expected to peak shortly after the onset of SN explosions, when the energy injection from SN clears low-density paths in the ISM \citep{Trebitsch17}. However, the coupling of SN feedback to the surrounding gas depends sensitively on local conditions such as density, metallicity, and the clustering of star-forming regions, leading to strong time variability in \fesc{} on short (few Myrs) timescales.

Despite the growing body of multi-wavelength studies on LCEs, the nature and relative importance of feedback mechanisms remain poorly understood. While SN feedback is widely invoked as the main driver of LyC leakage, observations of strong LCEs suggest that radiative feedback alone may sometimes be sufficient to ionize and clear the ISM possibly even before the first SN occur \citep{Komarova21, Amorin24, Carr25}. The balance between radiative and mechanical feedback and how it evolves with galaxy mass, metallicity, and star formation history is still an open question.

\section{Insights from radio studies using SKA precursors: JVLA, uGMRT and LOFAR}
\label{sec: ska precursors results}

Radio continuum (RC) observations with SKA precursors have recently provided independent insights on the conditions that enable LyC escape. However, these studies have only targeted relatively bright LCEs and a proper study of LCEs in the radio is still lacking. We summarise these recent results on LCEs at low-$z$ from three key SKA precursors: JVLA, uGMRT and LOFAR. Together, these works highlight how multi-frequency radio data offer a time-sensitive diagnostic of feedback, starburst age, and ISM conditions relevant for LyC photon escape. Next we briefly mention the various samples and details of the observations in the radio.

\citet{chakraborti2012} \revtext{targeted} the bright GP population using shallow GMRT 610 MHz data and from stacking the archival VLA images from the FIRST survey. The \gmrt{} study of BBs reaching a sensitivity of $\sim 10\mu$Jy at $\sim$1 GHz to probe their non-thermal emission and magnetic fields \citep{Sebastian19}. \citet{Borkar24-BB-LOFAR} studied a population of GPs and BBs using archival \lofar{} observations on GPs/BBs at $150$ MHz 
%found a substantial suppression in the radio relative to canonical SFR–radio (150 MHz) relations, and concluded that many GP/BB systems are under-luminous in radio for their optical/UV derived SFRs.  

The LzLCS \textit{JVLA} survey \citep{Bait24a} targeted 53 low-$z$ LyC-emitter candidates, obtaining deep S- and C-band ($2-8$ GHz) observations, with a subset also covered at L-band ($1-2$ GHz). Approximately half of the sources were detected at $\sim 5 - 10 \mu$Jy sensitivity, allowing robust measurements of radio spectral indices and continuum-based SFRs. 
The multi-frequency \rcsed{} work \citep{Bait25} combines \textit{uGMRT} (Band 5), \textit{JVLA} (S/C/X/Ku) and archival \textit{LOFAR} data to span $\sim 0.12 - 18$ GHz for a small sample of 8 extreme star-forming galaxies (xSFGs), enabling detailed modelling of spectral curvature and free–free absorption.

\begin{figure}[ht!]
    \centering
	\includegraphics[width=0.5\columnwidth]{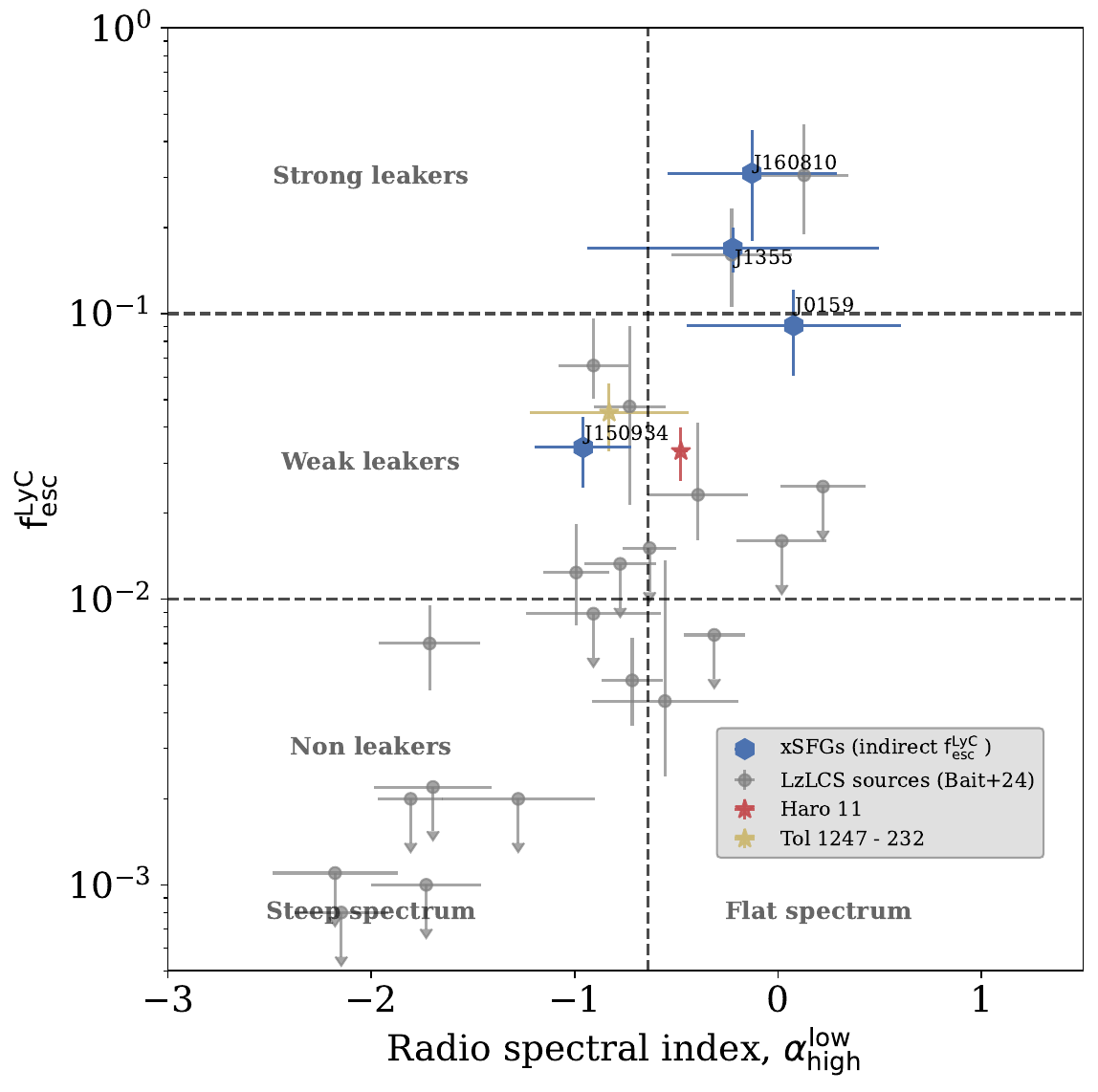}
\caption[Relation between \fesc{} and the radio spectral index]{\revtext{Relation between the Lyman continuum escape fraction (\fesc{}) and the integrated radio spectral index measured between the S~(3~GHz) and C~(6~GHz) bands. Galaxies with direct LyC observations from the LzLCS survey are plotted in gray points where the \fesc{} is derived using a UV-fit method \citep{Flury22a}, with non-detections represented as 1$\sigma$ upper limits. The blue hexagons represent xSFGs from \citet{Bait25} with indirect \fesc{} estimates. For reference, two nearby confirmed LyC-emitting galaxies, Haro~11 (red star) and Tol~1247--232 (yellow star), are also included. Horizontal dashed lines separate the regimes corresponding to strong (\fesc{}~$>$~0.1), weak (0.01~$<$~\fesc{}~$<$~0.1), and negligible (\fesc{}~$<$~0.01) escape fractions. The vertical dashed line marks the characteristic spectral index of a typical radio continuum spectrum (\alphacs{}~=~--0.64), distinguishing between flat- and steep-spectrum sources. See text for more discussion and the relevant section in \citet{Bait24a, Bait25} from where the figure is adopted.} }
    \label{fig: fesc radio spec index}
\end{figure}

%\subsection{Key observational results}
Below we briefly review the key results from these studies.
\begin{itemize}
    \item 
\textbf{Radio Spectral indices and RC-SFR offsets.}  
The LzLCS galaxies show a wide range of radio spectral indices between $3 - 6$ GHz, from flat ($\alpha \geq  -0.1$) to very steep ($\alpha \leq -1.0$ ), with a mean \alphacs{} $\sim -0.9$ and large scatter. Importantly, as seen in Figure \ref{fig: fesc radio spec index} the strongest LyC leakers preferentially exhibit \emph{flatter} spectra, whereas non-leakers tend to show steeper slopes. Radio‐derived SFRs are offset from canonical radio-SFR calibrations by $\sim$0.6 dex, though including the spectral index as a secondary parameter reduces the scatter in the SFR correlations.

\item \textbf{Suppressed non-thermal emission in GPs/BBs.}  
In the uGMRT study, \citet{Sebastian19} found that their radio‐based SFRs are suppressed compared to optical emission-line SFRs, a non-thermal fraction (median $\approx 0.49$) lower than for more evolved systems, and inferred equipartition magnetic fields which are significantly high for such compact dwarfs. They interpret this as either (i) a starburst so young that few SN have yet formed a steady CR/synchrotron population or (ii) efficient escape of CR electrons via galactic winds/outflows. \citet{Borkar24-BB-LOFAR} extended this to a larger bright GP/BB sample using archival \lofar{} data and show that a large fraction (~60–70 \%) of these low-mass, high‐sSFR systems fall systematically below the expected SFR–radio relation established for more massive, mature galaxies. These studies highlight that at the low-mass end, deviations from the canonical radio–SFR relation are significant, and that many GP/BB systems remain undetected or under‐luminous at radio frequencies, consistent with very young ages, inefficient CR build-up, or altered ISM/CR coupling.

\item \textbf{Thermal‐dominated SEDs and low‐frequency turnovers.}  
The multi-frequency \rcsed{} analysis from \citep{Bait25} of candidate LCEs  reveals flat spectra at high frequencies (6–15 GHz). In a few systems there is a significant low-frequency turnovers ($\sim$2–10 GHz) e.g., in SBS 0335-052 \citep{Hunt04} and in a few xSFGs from \citet{Bait25}. These SEDs are well described by thermally dominated models with strong free–free absorption, requiring very high emission measures and implying dense, compact H \textsc{ii} regions hosting $\leq$ 5 Myr‐old stellar populations.

\end{itemize}

%\subsection{Physical interpretation and common threads}

Across these studies, a consistent picture emerges. Flat GHz spectra and enhanced thermal fractions point to very young, compact starbursts producing abundant ionizing photons but not yet dominated by SN‐driven synchrotron emission; steeper spectra instead trace systems where SN and cosmic rays contributes a larger non-thermal component. The observed radio spectral index – \(f_{\rm esc}\) correlation therefore provides a time‐sensitive diagnostic of the starburst phase most conducive to LyC escape.

Low-frequency turnovers indicate either free–free absorption by dense ionised gas or deficits of aged CR electrons. High emission measures inferred from the thermal models suggest that star formation occurs in extremely dense, compact clusters. 

%The radio results from GP/BB samples (Borkar et al.\ 2024; Sebastian & Bait 2019) further show that even in very young systems, amplified magnetic fields (tens of µG) can be reached rapidly (via small‐scale dynamo), and efficient CR escape or lack of build‐up can suppress synchrotron emission. This suppression both alters the radio–SFR calibration and suggests that early feedback conditions (pre‐SN) may favour LyC escape, before the mature SN/CR feedback era begins.

%\subsection{Open questions and future directions}

However, several uncertainties remain. While the JVLA results suggest the radio spectral index–\(f_{\rm esc}\) correlation is largely orientation independent, \revtext{small‐scale ISM clumpiness can modify the amount of escaping LyC photons at the line of sight which can lead to a larger scatter in a larger population of LCEs.} The interpretation of low‐frequency turnovers remains degenerate between free–free absorption and synchrotron ageing, pointing to the need for spatially resolved, multi‐band radio SEDs combined with optical or mm‐wave diagnostics.

Because free–free and synchrotron components probe different timescales, e.g., ionising stars ($\leq 10$ Myr) versus \revtext{SN ($\geq 3 - 100$ Myr), wherein at earlier times the contribution from SN is sub-dominant.} Thus the radio spectral evolution offers constraints on the timing for feedback and escape processes (see Figure \ref{fig: fesc radio spec index}).

In summary, SKA‐precursor radio studies reveal a coherent, though nuanced, picture: flat, thermally dominated radio spectra and suppressed non-thermal emission are associated with strong LyC leakage, tracing extremely young, dense star clusters where radiative feedback may dominate. In contrast, steep, synchrotron‐dominated spectra mark more evolved starbursts where SN and CR feedback have matured, possibly reducing LyC escape. Multi-frequency radio diagnostics therefore provide a powerful, temporally sensitive window into the feedback processes governing LyC photon escape in galaxies.

\section{Detecting RC emission from LCEs and EELGs using SKA \aastar{} and AA4}
\label{sec: ska prediction}
In this section we discuss the prospects of detecting the fainter population of LCEs and EELGs using the SKA-mid frequencies using \aastar{} and AA4 which cannot be easily probed using the current generation radio telescopes.

\revtext{The RC emission at GHz frequencies} from a normal star-forming galaxy is composed of a combination of a flat ($\alpha = -0.1$) thermal (free-free) emission component and a generally steep non-thermal emission \citep{condon1992, murphy2011}. The thermal radio emission directly relates to the production rate of total (dust un-attenuated) LyC photons from \hii{} regions. The non-thermal arises due to synchrotron emission from relativistic electrons (also termed as CRs) accelerated by SN explosions under the presence of magnetic fields in the ISM. The non-thermal emission is directly related to the CR energy spectrum (or the non-thermal spectral index (\alphanth{})) and SN rate in a galaxy. By constraining the thermal/non-thermal spectrum (e.g., using multi-frequency radio observations) we can constrain the thermal radio fraction (\fth{}) and \alphanth{}. This allows us to independently constrain the SN rate in a galaxy using radio observations alone. 

Currently such a multi-frequency study of LCEs is lacking in the literature and thus the range of \fth{} and \alphanth{} spanned in these systems and its dependence on other physical parameters is unknown. Using a \vla{} 3- and 6-GHz study of LCEs a low-$z$ \citet{Bait24a} found that the apparent spectral index between these bands correlates with various other physical properties (e.g., \fesc{}, \Oratio{}, \mstar{} etc.), see also Sec. \ref{sec: ska precursors results}. However, a physical model relating these properties of LCEs in the radio is lacking. This \revtext{prevents} us to make predictions for the SKA using first principles. Nevertheless, we take a phenomenological approach to compare the radio and optical properties. 

Consider the radio spectrum of a normal star-forming galaxy, 

\begin{equation}\label{eq: simple sed model}
    S^{\mathrm{tot}}_{\nu} = S_{\nu}^{\mathrm{th}} + S_{\nu}^{\mathrm{nth}} 
    = S_{{1.4 \mathrm{GHz}}}^{\mathrm{th}}\left(\frac{\nu}{{1.4 \mathrm{GHz}}}\right)^{-0.1} + S_{{1.4 \mathrm{GHz}}}^{\mathrm{nth}}  \left(\frac{\nu}{{1.4 \mathrm{GHz}}}\right)^{\alpha_\mathrm{nth}}.
\end{equation}

We can then define the thermal fraction at an observed frequency ($\nu$) as, 

\begin{equation}\label{eq: thermal_fraction}
    \mathrm{f}_{\nu} ^{\mathrm{th} }  =  S_{\nu}^{\mathrm{th}} / S^{\mathrm{tot}}_{\nu}.
\end{equation}

Here the radio thermal flux density can be directly related to the dust-corrected H$\beta$ flux density (\FHbeta{}),  
\begin{equation} \label{eq: hunt_thermal_flux_hbeta_rel}
\left( \frac{\mathrm{F}^{\mathrm{dustcorr}}_{\mathrm{H}\beta}}{10^{-16}~\mathrm{erg~cm^{-2}~s^{-1}}} \right)
= 3.02 \times 
\frac{n(\mathrm{H}^+)}{n(\mathrm{H}^+) + n(\mathrm{He}^+)}~
\left( \frac{T_{e}}{10^4~\mathrm{K}} \right)^{-0.56}
\left( \frac{\nu}{\mathrm{GHz}} \right)^{0.1}
\left( \frac{S^{\mathrm{th}}_{\nu}}{\mu\mathrm{Jy}} \right).
\end{equation}
See \citet{Hunt04} for a derivation. Here $n(\mathrm{H}^+)$ and $n(\mathrm{He}^+)$ is the number density of ionized hydrogen and Helium, and $T_e$ is the electron temperature. In all our calculations we assume a $T_e$ of $20,000$ K generally observed in LCEs \citep[e.g.,][]{Izotov21b} and is higher than typically found in normal galaxies.  Following \citet{Hunt04}, we assume $n(\mathrm{He}^+)$/$n(\mathrm{H}^+)$ $\sim 0.08$ which is applicable for low-metallicity environments, typical of the LCE population. 

Using Eq. \ref{eq: thermal_fraction} and Eq. \ref{eq: hunt_thermal_flux_hbeta_rel}, we can relate \FHbeta{} and the total radio flux density for a fixed \fth{} as follows, 

\begin{equation} \label{eq: hbeta_radio_total_relation}
\left( \frac{S^{\mathrm{tot}}_{\nu}}{\mu\mathrm{Jy}} \right) = \left( \frac{0.358}{ \mathrm{f}_{\nu} ^{\mathrm{th} } }\right) \times \left( \frac{T_{e}}{10^4~\mathrm{K}} \right)^{0.56} \left( \frac{\nu}{\mathrm{GHz}} \right)^{-0.1} \left( \frac{\mathrm{F}^{\mathrm{dustcorr}}_{\mathrm{H}\beta}}{10^{-16}~\mathrm{erg~cm^{-2}~s^{-1}}} \right)
\end{equation}

\revtext{Note that in principle, a high \fesc{} suppresses the free--free emission, since ionizing photons that escape do not contribute to recombination within the H\,{\sc ii} regions. However, for the majority of known LCEs with \fesc{}~1--30\% this effect is minor, and becomes significant only for the most extreme leakers. Importantly, this does not affect the relation in Eq.~3, since both the thermal radio emission and the optical recombination lines trace the same absorbed ionizing photon budget, preserving their ratio.}

We can use this relation to predict the expected total radio flux density in various SKA observing bands using \FHbeta{} and \fth{}. Note that such a relation relates the two flux densities and are thus independent of the assumed cosmology and redshift of the source. Additionally, we do not choose to use SFR to predict the radio flux density as SFRs measured using relations at UV/optical/IR depends on various assumptions e.g., the initial mass function, star-formation history etc. On the other hand the relation in Eq. \ref{eq: hbeta_radio_total_relation} makes a comparison between two directly observed quantities. 

\begin{figure*}[ht!]
    \centering
	\includegraphics[width=1.0\columnwidth]{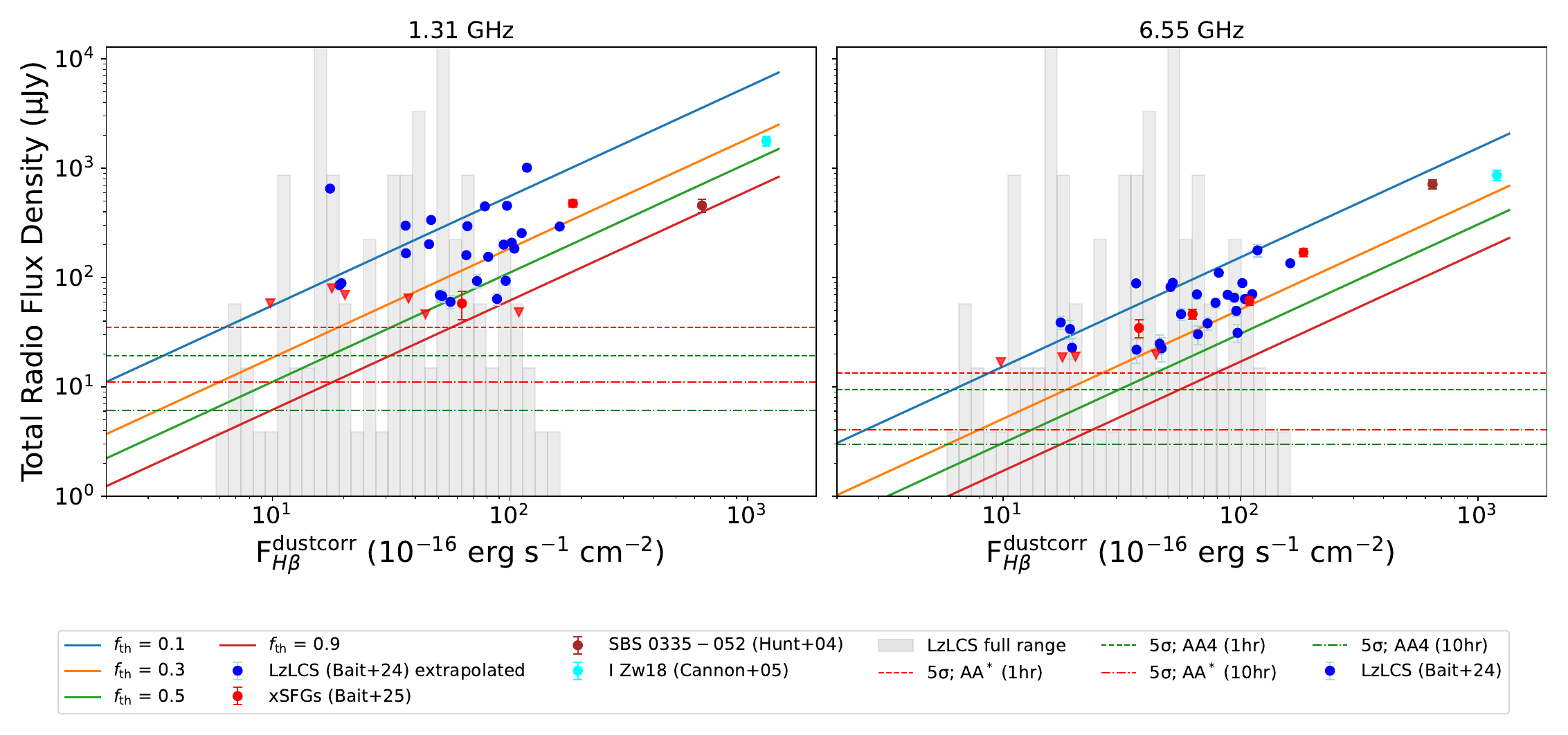 }
    \caption{The expected radio flux density at SKA-Mid Band 2 (left-panel) and Band 5a (right-panel) vs the dust-corrected H$\beta$ flux density. We show the relation from Eq. \ref{eq: hbeta_radio_total_relation} for different values of \fth{} in coloured solid lines. The dashed horizontal lines shows the 5$\sigma$ limit on the sensitivity of the two SKA-Mid bands assuming 1 hr on-source integration (see Table \ref{tab:SKA_RMS} for details). We show the radio data from SKA precursors (\vla{} and \gmrt{}) for the LzLCS sample (blue) at $z \sim 0.3$ from \citet{Bait24a} and xSFGs (red circles) at $z \sim 0$ from \citet{Bait25}, two well studied LCEs population in the radio, for reference. The red triangles denote 3$\sigma$ upper limits on xSFGs. In the background we show the full range of \FHbeta{} spanned by the LzLCS. We also show two well studied low-metallicity starbursts systems from the literature, SBS 0335-052 \citep{Hunt04} and I Zw18 \citep{Cannon05}. SKA-Mid bands observations with AA$*$ and AA4 allows us to detect the fainter LCEs and EELGs populations in the radio which cannot be easily probed using the current generation radio telescopes. }
    \label{fig: SFR Flux comparison}
\end{figure*}

In Figure \ref{fig: SFR Flux comparison} we plot this linear relation at two different SKA-Mid bands, Band 2 (left-panel) and Band 5a (right-panel). We plot for different values of \fth{} (at a reference frequency of $1$ GHz) from 0.1 to 0.9 in different colored solid lines. As expected for a given \FHbeta{}, a higher \fth{} leads to lower radio flux density due to a lower contribution of the non-thermal emission component and vice versa. Thus for a given \FHbeta{}, the solid lines somewhat shows the range of expected radio flux density due to varying \fth{}. In a few cases, it is possible that the radio flux density is below the $\fth{} = 1.0$ line, this is because in some cases the radio spectrum of LCEs shows a turnover at low-frequencies possibly due to free-free absorption \citep[FFA; see e.g.,][]{Bait25}, which is expected under high ionized gas density environment which is also typical of some LCEs. We show the 5$\sigma$ sensitivity limits of the two SKA-Mid bands for AA* (AA4) in red (green) dashed horizontal lines. We use the SKA sensitivity calculator \footnote{https://www.skao.int/en/science-users/ska-tools/493/ska-sensitivity-calculators} to estimate the RMS noise for different telescope setups. Here we assume \revtext{targeted} observations pointed on an individual LCE/EELG. In all the cases, we have assumed a tapering of $1.069$ leading to a synthesized beam close to $\sim 1''$ which is required as LCEs are known to be compact. We derive the RMS values for an on-source integration time of 1 and 10 hrs. Table \ref{tab:SKA_RMS} shows the expected RMS for the AA* and AA4 array for the above assumed observing setup. 

In Figure \ref{fig: SFR Flux comparison} we show the LzLCS observations using the \vla{} in blue points from \citep{Bait24a} for a reference. The \vla{} observations were originally performed in S- ($3.0$ GHz) and C- ($6.0$ GHz) bands. We have extrapolated the flux density using the observed spectral index to the SKA-Mid Band 2 (left-panel) and Band 5a (right-panel). We also show xSFGs from \citet{Bait25} which represent a sample of low-metallicity extreme starburst galaxies which are potential LCEs based on indirect indicators. Finally we also show two well-studied nearby and bright XMDs in radio from the literature, \sbs{} \citep{Hunt04} and \izw{} \citep{Cannon05} for reference..

\begin{table}[ht!]
\centering
\caption{Observing parameters for different SKA-Mid array configurations and observing bands.}
\label{tab:SKA_RMS}
\begin{tabular}{l l c c c c c}
\hline
\hline
Band & Array & Cent. Freq & BW & Int. time & Syn. beam & RMS \\
 &  & (GHz) & (GHz) & (hrs) & ($'' \times ''$) & ($\mu$Jy\,beam$^{-1}$) \\
\hline
\multirow{4}{*}{Mid band 2} & AA* &  \multirow{4}{*}{1.31} & \multirow{4}{*}{0.72} & 1  & 1.37 $\times$ 1.15 & 6.98 \\
& AA* &  &    & 10 & 1.37 $\times$ 1.15 & 2.21 \\
& AA4 &  &    & 1  & 1.09 $\times$ 1.07 & 3.86 \\
& AA4 &  &   & 10 & 1.09 $\times$ 1.07 & 1.22 \\
\hline
\multirow{4}{*}{Mid band 5a} & AA* &  \multirow{4}{*}{6.55} & \multirow{4}{*}{3.9} & 1  & 0.89 $\times$ 0.85 & 2.67 \\
& AA* &  &    & 10 & 0.89 $\times$ 0.85 & 0.81 \\
& AA4 &  &    & 1  & 0.88 $\times$ 0.86 & 1.89 \\
& AA4 &  &    & 10 & 0.88 $\times$ 0.86 & 0.596 \\
\hline
\end{tabular} \\
\small
\textbf{Notes.} RMS values are estimated using the official SKA sensitivity calculator where we have assumed uniform weighting and a tapering of 1.069$''$. 
\end{table}

From Figure~\ref{fig: SFR Flux comparison}, it is evident that the currently studied sample of LCEs with radio detections primarily represents the brighter end of the population, with \FHbeta{}~$\gtrsim 35 \times 10^{-16}$~erg s$^{-1}$ cm$^{-2}$. Only a few sources are detected below this threshold. We show the full distribution of LCEs from the LzLCS in grey histogram for a reference. This highlights a substantial population of fainter LCEs that remain unexplored in the radio. In the large sample of EELGs from \citet{Izotov21a-CSFGs} we can see that a large population of EELGs are below the above mentioned limit on \FHbeta{} and thus a general population of faint EELGs (GPs, BBs, XMDs etc.) also would remain unexplored in the radio with current SKA precursors. The two sources detected at \FHbeta{}~$\lesssim 35 \times 10^{-16}$~erg s$^{-1}$ cm$^{-2}$ exhibit relatively bright radio flux densities (corresponding to $\fth{} \gtrsim 0.1$). The rest of the xSFGs have non-detections in the radio, thus suggesting that LCEs and in general EELGs with fainter radio emission remain largely unexplored.

A large fraction of fainter LCEs, particularly those expected to follow the $\fth{} = 0.1$ (blue solid) relation, should be detectable with SKA-Mid AA$^*$ in just 1~hr of on-source integration. Deeper AA$^*$ observations (e.g., 10~hr) or observations with the more sensitive AA4 configuration will extend detections to fainter \FHbeta{} values and sources with higher thermal fractions. For instance, AA4 should detect most LCEs with $\fth{} = 0.3$ and nearly the entire population with $\fth{} = 0.5$ in 1~hr integrations. Detecting thermally dominated LCEs ($\fth{} = 0.9$) will be more challenging and likely require deeper, $\sim$10~hr AA4 observations. This thermally dominant faint population, though difficult to detect, is of particular interest as brighter LCEs with high thermal fractions tend to exhibit the strongest \fesc{} \citep[$\gtrsim 10\%$;][]{Bait24a, Bait25}.

\section{Detecting radio emission from LCEs and EELGs at high-$z$ using \aastar{} and AA4 observations and wide, deep and ultra-deep SKA-Mid surveys}
\label{sec: LCE high-z}
\revtext{The sections above have established both the physical motivation and the observational diagnostics for identifying LyC leakers and EELGs through their radio continuum emission. Here we turn to the question of how many such sources can realistically be detected with the SKA and its different array configurations and survey strategies. We derive the expected radio luminosity function (LF) of these populations from well-calibrated H$\alpha$ and EELGs LFs, then translate those luminosity distributions into predicted source counts and sky areas as a function of radio flux density at SKA-Mid Band~2 (1.31~GHz). Note that LCEs are among the population of H$\alpha$ emitters (HAEs) which have extreme emission-line properties, EELGs, which form a prime candidate for LCEs. Due to a lack of confirmed population of LCEs at high-$z$ (above $\sim 0.3$), we have to rely on a candidate LCE population.}

\subsection{The expected radio luminosity function of LCEs and EELGs at high-$z$}
\begin{figure*}[ht!]
    \centering
	\includegraphics[width=1.0\columnwidth]{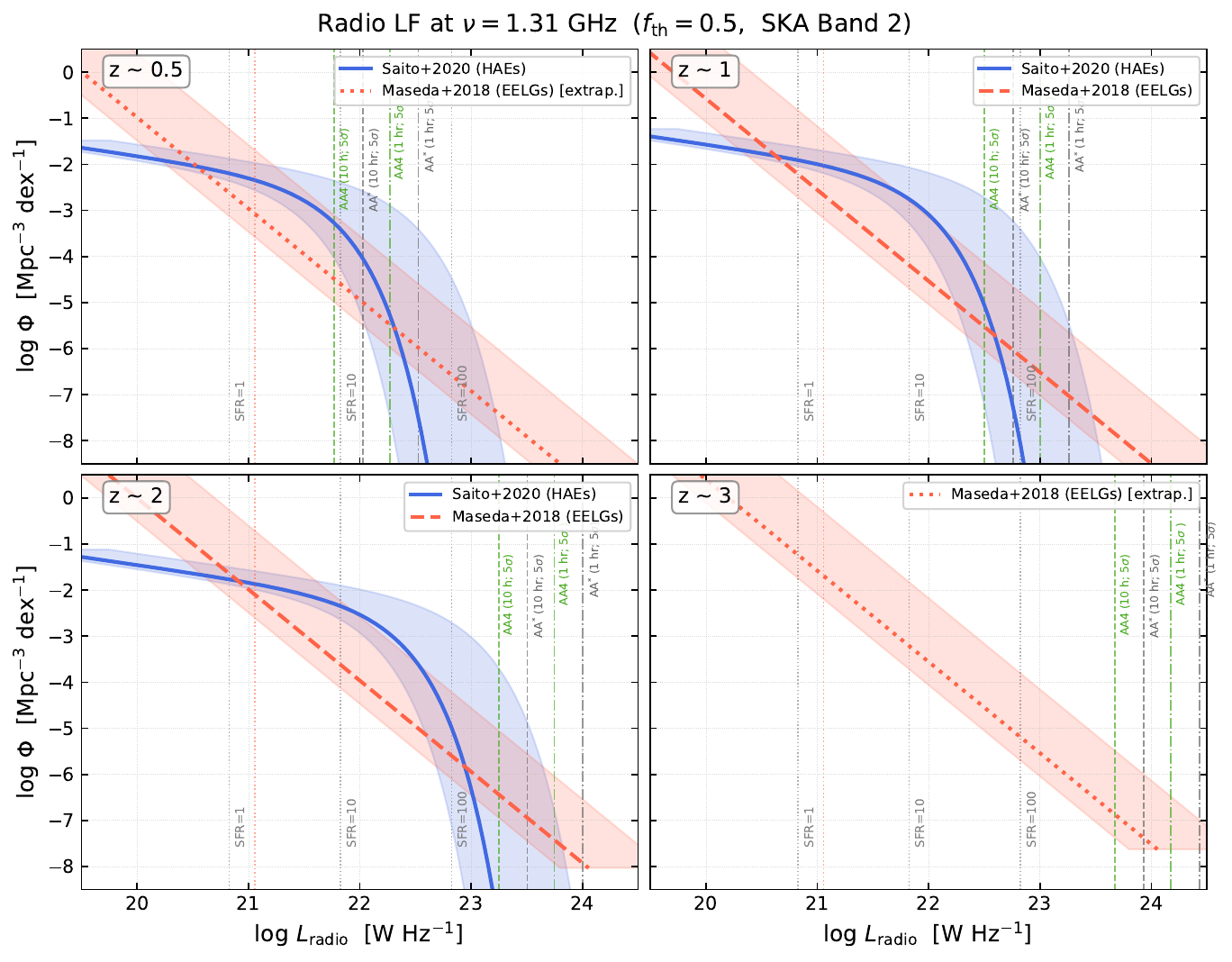}
    \caption{\revtext{Predicted radio spectral luminosity function of H$\alpha$ emitters and EELGs at $\nu = 1.31$~GHz (SKA-Mid Band~2) for $z = 0.5$, 1.0, 2.0, and 3.0, derived by applying Eq.~4 to two H$\alpha$ luminosity functions. Solid blue line shows the Schechter LF of \citet{Saito2020} for the general H$\alpha$-emitting galaxy population ($0.3 < z < 2.5$); dashed red line shows the power-law LF of \citet{Maseda2018} for extreme EELGs with EW$_{H\alpha} > 424$~\AA\ ($0.7 \lesssim z \lesssim 2.3$). Both LFs are evaluated at a fiducial thermal fraction \fth{} $= 0.5$, $T_e = 2\times10^4$~K, and $n(\mathrm{He}^+)/n(\mathrm{H}^+) = 0.08$. The shaded region shows the range of LF covered for \fth{} between 0.1 to 0.9. Vertical dashed and dash-dotted lines mark the $5\sigma$ flux density sensitivity of SKA-Mid Band~2 for \aastar{} (grey) and AA4 (green) in 1-hour and 10-hour integrations, at each redshift converted to a luminosity threshold. } }
    %The Saito LF shows the characteristic Schechter exponential suppression at high luminosities, while the Maseda EELG LF extends as a steep power law, highlighting the contribution of the most extreme star-forming systems to the high-luminosity end.}
    \label{fig:radio_LF}
\end{figure*}

\revtext{To predict the radio continuum LF of candidate LCEs and EELGs, we utilize the tight physical link between free--free (thermal) radio emission and the ionizing photon budget of H\,{\sc ii} regions, as encoded in Eq.~4 of this chapter (derived from \citealt{Hunt04}). That equation directly relates the dust-corrected H$\beta$ flux to the total radio flux density for a fixed thermal fraction \fth{}.
Because both $S_\mathrm{tot}$ and $F_{H\beta}$ scale identically with luminosity distance, the same coefficient connects the corresponding luminosities:}
\begin{equation}
    L_{\nu,\mathrm{radio}}\,[\mathrm{erg\,s^{-1}\,Hz^{-1}}] = \frac{0.358}{\fth{}}
    \left(\frac{T_e}{10^4\,\mathrm{K}}\right)^{0.56}
    \left(\frac{\nu}{\mathrm{GHz}}\right)^{-0.1}
    \frac{L_{H\alpha}}{2.86}\times 10^{-13},
\end{equation}
\revtext{where the intrinsic Case~B ratio $L_{H\alpha}/L_{H\beta} = 2.86$ is adopted, along with $T_e = 2\times10^4$~K and $n(\mathrm{He}^+)/n(\mathrm{H}^+) = 0.08$, values appropriate for low-metallicity, high-ionization \mbox{H\,{\sc ii}} regions representative of confirmed LCEs and \revtext{metal-poor galaxies} \citep[e.g.,][]{Izotov21b}.}

\revtext{We apply this conversion to two observationally motivated H$\alpha$ LFs. The first is the Schechter LF of \citet{Saito2020}, which parametrises the full population of H$\alpha$ emitters drawn from the EL-COSMOS survey over $0.3 < z < 2$, with a characteristic luminosity $\log(L^*_0/\mathrm{erg\,s^{-1}}) = 41.59$ and faint-end slope $\alpha = -1.35$, and both $\phi^*$ and $L^*$ allowed to evolve with redshift as power laws pivoting at $z_\mathrm{piv} = 1.53$. The second is the power-law LF of \citet{Maseda2018} for extreme emission-line galaxies (EELGs) selected from the 3D-HST survey in the CANDELS fields with rest-frame EW$_{[OIII]} > 500$~\AA{} and/or  EW$_{H\alpha} > 424$~\AA, derived from grism spectroscopy over $0.7 \lesssim z \lesssim 2.3$. They characterise the H$\alpha{}$ LF of EELGs using a single power-law given by a steep faint-end slope $\alpha = -1.98$ and a number density that evolves as $(1+z)^{3.25}$. The Maseda LF represents EELGs, which are the prime LCE candidates at high redshift. Together, the two LFs bracket the plausible range of the population of emission line galaxies at high-$z$ with the Saito LF represents the general H$\alpha$-emitting galaxy population providing an upper limit on the expected population, while the Maseda LF being close to the objects relevant to this chapter. }

\revtext{Figure~\ref{fig:radio_LF} shows the resulting radio spectral LF at $\nu = 1.31$~GHz for $z = 0.5$, 1, 2, and 3, evaluated at a fiducial \fth{} $= 0.5$. The Saito LF (blue solid line) exhibits the characteristic Schechter exponential cutoff at high luminosities ($L_{\nu} \gtrsim 10^{22}$~W~Hz$^{-1}$), reflecting the rarity of high-SFR systems. The Maseda EELG LF (red dashed line) follows a steeper power law with no bright-end cutoff in the probed luminosity range, which means it overtakes the Saito LF for extreme sources. Both LFs show positive redshift evolution over $z \sim 0.5$--$2$, tracking the rise in cosmic star formation rate density. Vertical lines indicate the $5\sigma$ flux density limits of the four \aastar{}/AA4 array--integration time combinations listed in Table~1; the corresponding luminosity at each redshift illustrates the depth of the radio LF that each configuration can probe. The most sensitive configuration, AA4 at 10~hours, reaches $5\sigma \simeq 6.1\,\mu$Jy at 1.31~GHz, probing radio luminosities $L_\nu \sim 10^{22}$--$10^{24}$~W~Hz$^{-1}$ across $z \sim 0.5$--$3$, corresponding to SFRs of a few to several tens to 100s of $M_\odot$~yr$^{-1}$. Such a deep integration should be able to capture the knee of the radio LF at the lowest redshift ($\sim 0.5$) and only the brighter ends at high-$z$. In all observing configurations the \revtext{faint end} of the radio LF cannot be probed using direct observations. \revtext{Possibly, gravitational lensing is the only available way to get an idea of the radio properties of this population at the faint end using SKA.} In the next section we describe the number density prediction and expected number counts for detections of a population of HAEs and EELGs in a SKA-Mid survey setup. }

\revtext{We note an important caveat regarding the luminosity range over which 
the Maseda LF is physically meaningful. Being parameterised as a pure 
power law, it lacks the exponential bright-end suppression of the 
Schechter function and, when extrapolated to high
SFR (bright EELGs) and low SFR (faint EELGs), predicts EELG 
number densities that exceed those of the general HAE population from 
\citet{Saito2020}. This is unphysical since EELGs constitute a 
sub-population of HAEs, and reflects the breakdown of the power-law 
extrapolation beyond the luminosity range directly probed by the 3D-HST 
grism survey \citep{Maseda2018}. We therefore restrict 
all integrations over the Maseda LF to the range 
$1 \leq \mathrm{SFR}/M_\odot~\mathrm{yr}^{-1} \leq 100$, within which 
the EELG number density is mostly below that of the general 
HAE population. Our predictions for EELGs should 
nonetheless be regarded as upper limits, with the true EELG counts 
potentially lying below our estimates.}

\subsection{Number density predictions of LCEs and EELGs in SKA-Mid survey}
\label{sec:number_density}
\begin{figure*}[ht!]
    \centering
    \includegraphics[width=0.7\columnwidth]{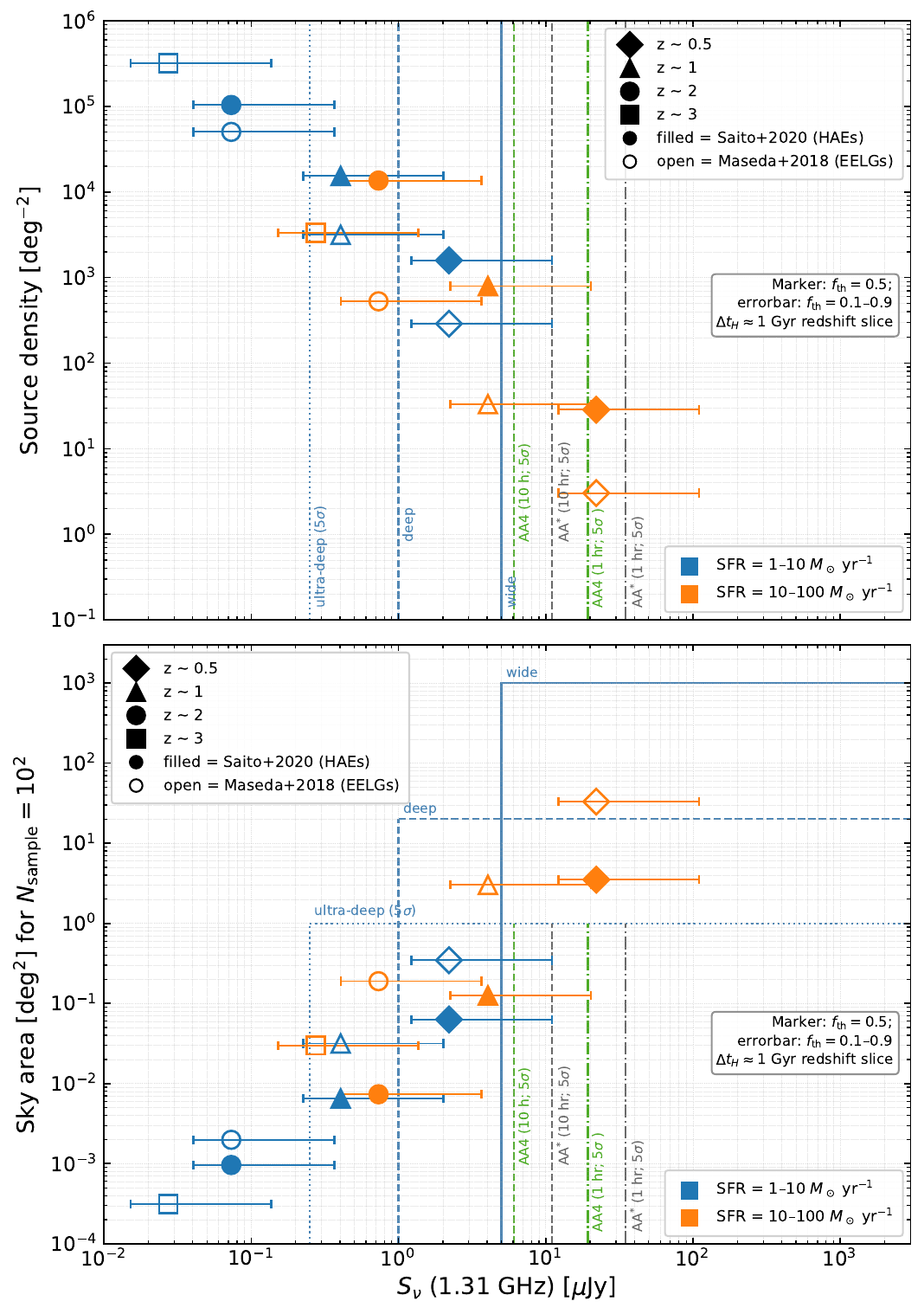}
    \caption{\revtext{Predicted surface number density (top) and required survey sky area (bottom) of H$\alpha$-emitting LCE candidates and EELG as a function of radio flux density $S_\nu$ at 1.31~GHz (SKA-Mid Band~2). Data points show predictions for four redshifts ($z \sim 0.5$: diamonds; $z \sim 1$: triangles; $z \sim 2$: circles; $z \sim 3$: squares) and two SFR bins (blue: $1$--$10\,M_\odot\,\mathrm{yr}^{-1}$; orange: $10$--$100\,M_\odot\,\mathrm{yr}^{-1}$), with horizontal error bars spanning the range \fth{} $= 0.1$--$0.9$. Filled symbols use the \citet{Saito2020} Schechter LF for all H$\alpha$ emitters; open symbols use the \citet{Maseda2018} power-law LF for extreme EELGs (EW$_{H\alpha} > 424$~\AA). The x-axis position of each point corresponds to the fiducial \fth{} $= 0.5$ prediction, converted from SFR to $L_{H\alpha}$ and then to $S_\nu$ using Eq.~4. Vertical grey (dash-dotted and dashed) and green lines mark the $5\sigma$ detection thresholds of \aastar{} and AA4 in 1-hour and 10-hour integrations, respectively; blue lines mark the extragalactic continuum reference survey scenario of an ultra-deep (${\sim}0.25\,\mu$Jy), deep (${\sim}1\,\mu$Jy), and wide (${\sim}5\,\mu$Jy) SKA-Mid survey \revtext{$5\sigma$ sensitivities} from \citet{Prandoni01.2026.SKA}. The bottom panel shows the sky area for 100 sources; horizontal reference lines illustrate the sky area of ultra-deep, medium, and wide surveys.}}
    \label{fig:density_skyarea}
\end{figure*}

\revtext{We use the radio LF predictions to compute the expected surface number density of detectable LCE and EELG candidates as a function of radio flux density $S_\nu$ at 1.31~GHz, and the corresponding sky area required to detect a population of 100 sources. These predictions directly inform the design of SKA-Mid survey strategies for high-redshift LyC-leaker searches. For our current analysis we use the SKA-Mid extragalactic continuum reference survey scenarios described in detail in another chapter \citep{Prandoni01.2026.SKA}.}

\revtext{For each H$\alpha$ LF and each redshift, we integrate the LF between the luminosity limits corresponding to each SFR bin converted from SFR via the \citet{murphy2011} calibration, between H$\alpha$ and SFR, over the comoving volume in a redshift slice $\Delta z$ corresponding to a lookback-time interval of $\Delta t_H \approx 1$~Gyr centred on the target redshift.
We use two SFR bins: low-SFR ($1$--$10\,M_\odot$~yr$^{-1}$) and high-SFR ($10$--$100\,M_\odot$~yr$^{-1}$) which are converted to the corresponding H$\alpha$ luminosity. As described in the previous section we then covert these H$\alpha$ luminosities to radio luminosities. In Figure~\ref{fig:density_skyarea} the x-axis position of each data point on the number density/sky area vs. radio flux density plot shows the predicted radio flux density at the midpoint of the SFR bin using the fiducial \fth{} $= 0.5$, while the horizontal error bars span the range from \fth{} $= 0.1$ (non-thermal dominated: brighter in radio for a fixed $L_{H\alpha}$) to \fth{} $= 0.9$ (thermally dominated: fainter in radio). This range captures the full uncertainty in the radio-to-$L_{H\alpha}$ conversion expected for the diverse LCE and EELG population. }

\revtext{Figure~\ref{fig:density_skyarea} (top panel) shows the predicted surface density as a function of $S_\nu$ at 1.31~GHz for sources at $z \sim 0.5$, 1, 2, and 3 (diamond, triangle, circle, and square symbols, respectively), spanning SFR bins colour-coded in blue ($1$--$10\,M_\odot\,\mathrm{yr}^{-1}$) and orange ($10$--$100\,M_\odot\,\mathrm{yr}^{-1}$). Filled markers correspond to the \citet{Saito2020} Schechter LF for all H$\alpha$ emitters (HAEs); open markers correspond to the \citet{Maseda2018} power-law LF for EELGs. At $z \sim 1$--$2$, where both LFs are simultaneously valid and the cosmic SFR density peaks, the HAEs LF predicts surface densities of $\sim 10^{3}$--$10^{5}$~deg$^{-2}$ across the SFR range probed, while the Maseda EELG LF gives surface densities $\sim 1$--$2$~orders of magnitude lower, reflecting the intrinsic rarity of the most extreme EELGs. }
%The difference between the two populations narrows at brighter flux densities (higher SFR or lower $z$), where the power-law EELG LF contributes more significantly relative to the exponentially suppressed tail of the Schechter function.

\revtext{Vertical lines in Figure~\ref{fig:density_skyarea} mark the flux density limits of the three SKA-Mid   extragalactic continuum reference survey scenario (ultra-deep: $\sim 0.25\,\mu$Jy, deep: $\sim 1.0\,\mu$Jy, wide: $\sim 5.0\,\mu$Jy at $5\sigma$; blue lines) described in \citet{Prandoni01.2026.SKA} and the four \aastar{}/AA4 configuration limits from Table~1 (grey for \aastar{}, green for AA4). The most sensitive pointed configurations—AA4 at 10~hr ($5\sigma \simeq 6.1\,\mu$Jy) and \aastar{} at 10~hr ($5\sigma \simeq 11.1\,\mu$Jy)—along with the SKA-Mid wide area survey can detect H$\alpha$ emitters and EELGs with SFR~$\gtrsim 10\,M_\odot$~yr$^{-1}$ at $z \sim 0.5$ with surface densities of order $\sim$several deg$^{-2}$. These wide area surveys can also start to probe the population of high-SFR and high \fth{} HAEs and EELGs at $z \sim 1$ in the range of $\sim~30 - 1000$ sources deg$^{-2}$ which can be sufficient to accumulate meaningful samples across a $\sim 10$~deg$^2$ deep survey. The deep and ultra-deep survey ($0.25\,\mu$Jy at $5\sigma$), unlocks access to low-SFR systems at $z \sim 0.5$--$1$ with densities $\gtrsim 100$~deg$^{-2}$, enabling statistical studies of the faint EELG and LCE population. }

\revtext{For context, we compare our predicted HAE and EELG number densities with those 
of the general star-forming galaxy (SFG) population \citep[see Fig.~3 in ][]{Prandoni01.2026.SKA}. We observe that our number counts on HAEs agree well with the general population of SFGs at the brighter end (close to $S_\nu \approx 10 \mu$Jy). And the EELGs number counts are below both SFGs and HAEs confirming the general rarity of these sources. At the fainter end we predict much higher number of both HAEs and EELGs compared to SFGs. This difference is driven by the generally steeper faint end slope of the H$\alpha$ LF of HAEs \citep{Saito2020} and EELGs \citep{Maseda2018} compared to the one used in \citet{Prandoni01.2026.SKA} from the T-RECS simulations \citep{Bonaldi19}. }

\revtext{The bottom panel of Figure~\ref{fig:density_skyarea} presents the inverse quantity: the sky area required to detect at least 100 sources as a function of the radio flux density. We find that wide (deep) area survey will be crucial to study a statistically relevant population of HAEs and EELGs with high-SFR (low-SFR) between $z \sim 0.5 - 1$. Ultra-deep surveys will be crucial to study the population of HAEs and EELGs at high-SFR at $z \geq 1$ in a single pointing. While such a survey can also capture the low-SFR population of HAEs and EELGs at $z \sim 1$, low-SFR population at $z \sim 2-3$ will be difficult to detect except in the case of sources with a dominant non-thermal emission component (high \fth{}). }

\revtext{Overall, these predictions based on HAEs and EELGs suggest that SKA-Mid offers a realistic pathway to statistical detections of high-redshift LCEs across a wide range of survey configurations. Moderate-depth surveys ($\sim$few $\mu$Jy, $\sim$20~deg$^2$) with AA4 or \aastar{} can already yield samples of $\sim$tens of H$\alpha$-bright EELGs at $z \sim 1$--$2$, while ultra-deep integrations will push into the low-SFR (and thus low stellar mass), high-redshift regime of LCEs—the most likely analogs of reionization-era leakers—are expected to reside. These number count predictions, combined with the radio spectral diagnostics described in the earlier sections, underscore the unique power of the SKA to characterise and ultimately identify the population of galaxies driving cosmic reionization. Finally, we caution that our number density predictions are based on simplistic assumptions on the intrinsic radio-SED and relies mainly on a varying \fth{} to incorporate the diverse radio properties. Local multi-frequency radio studies of LCEs show a great diversity in their radio spectral properties which is dependent on other physical parameters \citep[e.g., \Oratio{}, stellar mass, \fesc{} etc; see ][]{Bait24a} which can lead to additional uncertainity in our number density estimates presented here. }

\revtext{Finally, we note that the number count predictions presented in 
here assume that a sufficiently large 
sample of LCE and EELG targets are available for SKA-Mid follow-up. 
At present, the sample of spectroscopically confirmed LCEs remains 
modest--- $\sim100$ objects at low redshift from targeted 
HST observations \citep[e.g., LzLCS;][]{Flury22a}, and a small set of direct 
detections at $z \sim 1$--$2$ from AstroSat/UVIT deep fields 
\citep[e.g.,][]{Saha20, Maulick25}. This sample size may itself 
form the limiting bottleneck for future SKA-Mid studies. However, several upcoming 
facilities will transform this landscape. JWST is already 
delivering large samples of EELG candidates at $z \sim 1$--$6$ 
selected via indirect LyC diagnostics such as high \Oratio{} and 
large H$\beta$ equivalent widths \citep[e.g.,][]{Withers23_JWST_EELG, Davis_JWST_EELG, Boyett_JWST_EELG, Llerena_JWST_EELG}. The \textit{Nancy Grace Roman 
Space Telescope} will provide slitless near-infrared spectra for 
millions of H$\alpha$ emitters at $z \sim 1$--$2$ across thousands 
of square degrees \citep{Roman_HLSS_Wang22}, enabling statistical pre-selection of high-priority 
SKA-Mid targets. Further, the \textit{Ultraviolet Explorer} 
\citep[UVEX;][]{Kulkarni21}, a NASA proposed mission planned for 
launch $\sim$2030, will survey the sky in the far- and 
near-UV with sensitivity more than $50/100\times$ better than GALEX, 
enabling direct LyC detection in low-redshift leakers over large 
sky areas. Taken together, these facilities will substantially 
enlarge the confirmed LCE and EELG census, ensuring that the 
statistical power of future SKA-Mid surveys is not limited by 
the availability of known targets.}

\section{Conclusions}
\label{sec: conclusions}
The advent of the SKA will mark a major step forward in understanding the physical mechanisms that regulate the escape of LyC photons from galaxies.  
Building upon recent radio studies of low-redshift LyC emitters and metal-poor starbursts using SKA precursors, it is becoming increasingly clear that radio continuum diagnostics, particularly the spectral index and thermal fraction, provide unique, time-sensitive probes of stellar feedback, ionization conditions, and cosmic-ray processes.  
These findings indicate that radio observations can serve as powerful tracers of the feedback pathways that influence LyC leakage.
 
The increased sensitivity and frequency coverage of SKA-Mid, especially with the AA$^*$ and AA4 configurations, will allow detections of LyC emitters well below the flux limits of current facilities.  
As shown in this work, integrations of 1--10 hours with SKA-Mid Bands~2 and~5a will probe the fainter LyC-emitting population, and generally the population of faint EELGs (GPs, BBs, XMDs etc) at low-$z$ ($z \leq 0.3$) which are currently unexplored in the radio thus extending beyond the currently known bright systems with $\FHbeta{} \gtrsim 3\times10^{-15}$~erg~s$^{-1}$~cm$^{-2}$.  
This will enable investigations into the diversity of radio thermal fractions and radio spectral indices, linking them directly to ionizing photon escape fractions and interstellar medium conditions.  
Such observations will also mitigate existing selection biases, providing a more complete view of starburst phases that are thermally dominated or cosmic-ray deficient, and therefore potentially most conducive to LyC escape. 
 
The number density predictions presented in Section~\ref{sec: LCE high-z} quantify this potential in concrete terms. Using the H$\alpha$ LF of \citet{Saito2020} as an upper limit for the general HAE population, and the EELG-specific power-law LF of \citet{Maseda2018} as the more directly relevant tracer of LCE candidates, we find that the Maseda EELG surface densities are $\sim$1--2 orders of magnitude below those of the general HAE population at $z \sim 1$--$2$, reflecting the intrinsic rarity of the most extreme star-forming systems. Nevertheless, SKA-Mid wide-area surveys ($\sim$5\,$\mu$Jy at $5\sigma$, $\gtrsim$1000~deg$^2$) can access high-SFR ($\gtrsim 10\,M_\odot$~yr$^{-1}$) EELGs at $z \sim 1$, while moderately deep surveys ($\sim$1\,$\mu$Jy at $5\sigma$, $\sim$20~deg$^2$) with AA4 or \aastar{} can yield samples of tens to a few hundreds of H$\alpha$-bright EELGs at $z \sim 1$--$2$. Ultra-deep integrations ($\sim$0.25\,$\mu$Jy at $5\sigma$) are required to access the low-SFR ($1$--$10\,M_\odot$~yr$^{-1}$) regime at $z \gtrsim 1$, which likely hosts the closest analogs to the faint, reionization-era LCEs. The bottom panel of Figure~\ref{fig:density_skyarea} further shows that a sky area of $\sim$100~deg$^2$ or less is sufficient to assemble a sample of 100 high-SFR EELGs at $z \sim 0.5$--$1$ in a deep survey, making such a programme feasible within a single SKA Large Programme. Ultra-deep SKA surveys should assemble a population of $\sim 100$ HAEs and EELG sources in a single pointing at redshift of $0.5 - 1$ ($z\sim 2 - 3$) at the low-SFR (high-SFR) regime. \revtext{We emphasise that the EELGs detectable in such surveys are 
\textit{LCE candidates} selected by their extreme star-forming 
properties, not confirmed LyC leakers: they represent the galaxy 
population among which LCEs are expected to be found, but 
confirming the escape of ionizing radiation in individual sources 
will require complementary observations, such as direct 
LyC imaging at $z \lesssim 0.3$ with UV-sensitive observatories, 
or indirect diagnostics (\Oratio{}, Ly$\alpha$ morphology, \fesc{} 
proxies) at higher redshifts with JWST or the Roman Space Telescope.} Additionally, we caution that these predictions rest on the simplified assumption of a fixed radio SED shape parametrised solely by \fth{}, whereas local multi-frequency studies of LCEs reveal a rich diversity in radio spectral properties \citep{Bait24a, Bait25} driven by ionization parameter, stellar mass, and LyC escape fraction \citep[e.g.,][]{Izotov20}, implying that the true scatter in predicted source counts may be larger than the \fth{} uncertainty alone. 
 
Ultimately, such observations with SKA will help bridge detailed radio studies of high-z LCEs and the general population of low-$z$ analogues with observations of the faint, reionization-era galaxies being uncovered by JWST.  
Through its sensitivity and resolution, the SKA will connect the physics of radio continuum emission by probing supernova feedback, cosmic-ray population, and magnetic field to the global process of cosmic reionization, providing the missing physical insights on the first galaxies in the Universe.

\section*{Acknowledgments}
We thank the referee for their helpful comments which improved this chapter. Omkar Bait was supported by the National Science Foundation under Cooperative Agreement 2421782 and the Simons Foundation grant MPS-AI-00010515 awarded to the NSF-Simons AI Institute for Cosmic Origins — CosmicAI, https://www.cosmicai.org/.
%\section{Synergy between ALMA and SKA MID to study the radio-IR correlation in LCEs and XMDs: at low- and high-$z$}

%\section{Possible synergies with ELTs, JWST for a multi-wavelength study}

%\subsection{Figures}

%\begin{figure}[h]
%    \centering
%	\includegraphics[width=0.3\columnwidth]{SKAO Pictorial mark-01.png}
%    \caption{Example figure. Place caption below table.}
%    \label{fig:example_figure}
%\end{figure}

%Referring to Fig.~\ref{fig:example_figure}.

\bibliographystyle{abbrvnat-maxbibnames4}
\bibliography{chapter} % if your bibtex file is called example.bib

\end{document}